\documentclass{PoS}

\usepackage{bm}
\usepackage{amsmath,amssymb,slashed,amsfonts,latexsym,bbold}
\newcommand{\be}{\begin{equation}}
\newcommand{\ee}{\end{equation}}
\newcommand{\ba}{\begin{eqnarray}}
\newcommand{\ea}{\end{eqnarray}}
\newcommand{\la}{\langle}
\newcommand{\ra}{\rangle}
\newcommand{\di}{\mathrm{d}}

\title{Chiral-odd GPDs in large--$N_c$ QCD}

\ShortTitle{Chiral-odd GPDs in large--$N_c$ QCD}

\author{
  \speaker{P.~Schweitzer} \\ 
  Department of Physics, University of Connecticut, Storrs, CT 06269, USA\\
  E-mail: \email{peter.schweitzer@phys.uconn.edu}}

\author{
  C.~Weiss\\
  Theory Center, Jefferson Lab, Newport News, VA 23606, USA\\
  E-mail: \email{weiss@jlab.org}}

\abstract{We study the flavor structure of the nucleon's chiral-odd generalized 
parton distributions (transversity GPDs) in the large--$N_c$ limit of QCD. 
It is found that in the distributions $H_T$ and $\tilde E_T$ the flavor--nonsinglet 
component $u - d$ is leading in the $1/N_c$ expansion, while in $E_T$ and $\tilde H_T$ 
it is the flavor--singlet component $u + d$. This pattern is consistent with the 
flavor structure extracted from hard exclusive $\pi^0$ and $\eta$ electroproduction data, 
assuming that the processes are dominated by the twist--3 mechanism 
involving the chiral-odd pseudoscalar meson distribution amplitudes.}

\FullConference{QCD Evolution 2015 -QCDEV2015-\\
		26-30 May 2015\\
		Jefferson Lab (JLAB), Newport News Virginia, USA}

\begin{document}

\section{Introduction}
\label{Sec-1:introduction}

Generalized parton distributions (or GPDs) unify the concepts of parton density and
elastic form factor and enable a comprehensive description of the nucleon's quark 
and gluon single--particle structure in QCD; see 
Refs.~\cite{Goeke:2001tz,Diehl:2003ny,Belitsky:2005qn,Boffi:2007yc} for a review. 
At the leading--twist level the nucleon's
quark structure is described by 4 chiral-even (quark helicity--conserving) and 4 
chiral-odd (quark helicity--flipping) GPDs \cite{Diehl:2001pm}. The chiral-even
GPDs reduce to the usual unpolarized and polarized quark parton distribution functions
(PDFs) in the limit of zero momentum transfer. These GPDs appear in the amplitudes
of hard exclusive processes such as deeply virtual Compton scattering 
and exclusive meson production with longitudinal photon polarization,
for which QCD factorization theorems have been derived, and can be accessed experimentally 
in this way \cite{Ji:1996nm,Radyushkin:1997ki,Belitsky:2001ns,Collins:1996fb}.
The chiral-odd GPDs reduce to the quark transversity PDFs in the limit of 
zero momentum transfer. Relating these GPDs to hard exclusive processes has proved 
to be more challenging, as they decouple from vector meson production at leading twist 
in all orders in perturbative QCD due to the chirality requirements for massless 
fermions \cite{Collins:1999un}. The chiral-odd GPDs have been associated with the diffractive 
electroproduction of two vector mesons at leading twist \cite{Ivanov:2002jj},
but the process is difficult to measure \cite{Enberg:2006he,Beiyad:2010cxa,Pire:2014fwa}, 
and no data are available at present. Recent work suggests that pseudoscalar meson 
electroproduction ($\pi^0, \eta, \pi^+$) may be dominated by a twist--3 mechanism
involving the chiral-odd nucleon GPDs and the chiral-odd distribution amplitude, 
which originates from the dynamical breaking of chiral symmetry 
in QCD \cite{Ahmad:2008hp,Goloskokov:2009ia,%
Goloskokov:2011rd,Goldstein:2012az}. While no factorization 
theorem exists at this level, the pseudoscalar production amplitudes have been calculated 
in the modified hard scattering approach of Ref.~\cite{Goloskokov:2009ia,Goloskokov:2011rd}, 
which implements suppression of large--size $q\bar q$ configurations in the meson
through the QCD Sudakov form factor. The results agree well with the
$\pi^0$ and $\eta$ electroproduction data from the JLab CLAS experiment at 6 GeV 
incident energy, regarding both the absolute cross sections and the $L/T$ ratio
as inferred from the azimuthal--angle--dependent response functions
\cite{Bedlinskiy:2012be,Kubarovsky:2014spin}. This has opened the prospect of 
more detailed studies of the chiral-odd GPDs in further measurements 
of pseudoscalar meson electroproduction with the JLab 12 GeV Upgrade.

To further explore this possibility, it is necessary to gain more insight into the 
properties of the chiral-odd GPDs and their relation to other measures of nucleon 
structure. Contrary to the chiral-even GPDs, in the chiral-odd case neither the 
zero-momentum transfer limit of the GPDs (transversity PDFs) nor the local operator 
limit of the GPD (form factor of local tensor operator) correspond to structures that 
are easily measurable, so that little useful information can be obtained in this way.
The transversity PDFs can be extracted from observables in semi-inclusive deep-inelastic 
scattering; see Refs.~\cite{Bacchetta:2012ty,Anselmino:2013vqa,Kang:2014zza}
and references therein. 
The form factors of local tensor operators, which constrain the lowest $x$--moment of 
the chiral-odd GPDs, have been calculated in lattice QCD \cite{Gockeler:2006zu}
and in the chiral quark-soliton model \cite{Kim:1995bq,Ledwig:2010tu,Ledwig:2011qw}.
The $x$--dependent chiral-odd GPDs have been studied in quark bound--state models 
of nucleon structure \cite{Pasquini:2005dk,Burkardt:2007xm,Chakrabarti:2008mw}.
Besides these estimates not much is known about the properties of the chiral-odd GPDs.

Here we report about a study of the flavor structure of the chiral-odd GPDs in QCD in
the limit of a large number of colors (large--$N_c$ limit) \cite{inprep}. The large--$N_c$
limit represents a rigorous approach to QCD in the non-perturbative domain and leads to
model-independent relations governing meson and nucleon structure at hadronic
energies \cite{'tHooft:1973jz}. In the large-$N_c$ limit QCD becomes semi-classical,
and baryons can be described by mean field solutions in terms of meson 
fields \cite{Witten:1979kh}. It is assumed that the qualitative properties of QCD, 
such as the dynamical breaking of chiral symmetry in the ground state, are 
preserved as the limit is taken \cite{Coleman:1980mx}. 
While the dynamics at large $N_c$ remains complex 
and cannot be solved exactly, and the form of the mean field solution is not known, 
important insights can be obtained by exploiting known symmetry properties of the 
mean field \cite{Witten:1983tx,Balachandran:1982cb}. 
In this way one obtains relations for baryon mass 
splittings, meson-baryon coupling constants, electromagnetic 
and axial form factors, and other observables, which are generally 
in good agreement with observations, see Ref.~\cite{Jenkins:1998wy} for a review.
In the matrix elements of quark bilinear operators (vector or axial 
vector currents, tensor operators) the large--$N_c$ limit identifies leading and
subleading spin--flavor components. The approach can be extended to parton 
densities \cite{Diakonov:1996sr,Efremov:2000ar}, where e.g.\ it suggests a large 
flavor asymmetry of the polarized antiquark distribution $\Delta\bar u - \Delta\bar d$ 
as seems to be supported by the recent RHIC $W^\mp$ production 
data \cite{Aggarwal:2010vc,Adare:2010xa}.

Our study generalizes previous results on large--$N_c$ relations for chiral-even GPDs 
\cite{Goeke:2001tz}, quark transversity 
distributions \cite{Pobylitsa:1996rs,Schweitzer:2001sr,Pobylitsa:2003ty},
and 
matrix elements of local chiral-odd operators
\cite{Kim:1995bq,Ledwig:2010tu,Ledwig:2011qw} and relies on the formal apparatus
developed in these earlier works (see Sec.3 of Ref.~\cite{Goeke:2001tz}). 
In this note we review the basic properties of chiral-odd GPDs (Sec.~\ref{Sec-2:GPDs}),
describe the $1/N_c$ expansion of the chiral-odd GPDs (Sec.~\ref{Sec-3:large-Nc}),
and compare the pattern with the pseudoscalar meson electroproduction data
(Sec.~\ref{Sec-5:comparison}).

\section{Chiral-odd GPDs}
\label{Sec-2:GPDs}

GPDs parametrize the non-forward nucleon matrix elements of QCD light--ray operators
of the general form \cite{Goeke:2001tz,Diehl:2003ny,Belitsky:2005qn,Boffi:2007yc}
\be\label{Eq:generic-matrix-element}
	{\cal M }(\Gamma) = 
	P^+\int\frac{\di z^-}{2\pi}\,e^{ixP^+z^-}  
	\la N(p^\prime,\lambda^\prime)|\psi(-\tfrac{z}{2})\;\Gamma\,
	\psi(\tfrac{z}{2})\,|N(p,\lambda)\ra\biggl|_{z^+=0,\;\vec{z}_T=0} ,
\ee
where $P \equiv \frac{1}{2}(p' + p)$ is the average nucleon 4--momentum, $z$ is the displacement
of the quark fields, and the 4--vectors are described by their light-cone components 
$z^\pm = (z^0 \pm z^3)/\sqrt{2}, \, \vec{z}_T=(z^1,z^2),$ etc. $\Gamma$ denotes a generic
matrix in spinor and flavor indices and defines the spin--flavor quantum numbers of 
the operator. In the chiral-odd case the spinor matrix is of the form $\Gamma = i\sigma^{+j}$,
and the matrix element is parametrized as
\cite{Diehl:2001pm}
\ba\label{Eq:def-chiral-odd-GPD}
	{\cal M}(i\sigma^{+j}T^q)
	&=&
	\bar{u}(p^\prime,\lambda^\prime)\biggl[
	i\sigma^{+j}\;H_T^q
	+\frac{P^+\Delta^j-\Delta^+P^j}{M_N^2}\;\tilde{H}_T^q\nonumber\\
	&& 
	+\frac{\gamma^+\Delta^j-\Delta^+\gamma^j}{2M_N}\;E_T^q 
	+\frac{\gamma^+P^j-P^+\gamma^j}{M_N}\;\tilde{E}_T^q\biggr]
	u(p,\lambda) \,.
\ea
The GPDs $H_T^q=H_T^q(x,\xi,t)$, etc., are functions of the quark plus momentum 
fraction $x$, the plus momentum transfer $\xi = -\Delta^+/(2 P^+) = 
(p-p^\prime)^+/(p+p^\prime)^+$, and the invariant momentum transfer $t = \Delta^2 = (p' - p)^2$.
(For brevity we do not indicate the dependence of the GPDs on the normalization scale.)
Here we consider GPDs corresponding to a given quark flavor $q = (u, d)$, and $T^q$ 
denotes the corresponding projector on quark flavor indices.

As their chiral-even counterparts, the chiral-odd GPDs satisfy certain symmetry relations 
in $\xi$ due to time reversal invariance,
\be\label{Eq:xi-dependence}
	{\rm GPD}(x,-\xi,t) = \begin{cases}
     +\,{\rm GPD}(x,\xi,t)&{\rm for}\;\;{\rm GPD}=H_T^q,\;\tilde{H}_T^q,\;E_T^q ,\\
     -\,{\rm GPD}(x,\xi,t)&{\rm for}\;\;{\rm GPD}=\tilde{E}_T^q .\end{cases}
\ee 
Their integrals over $x$ (or first moments) 
coincide with the form factors of the local tensor operator
$\bar{\psi}(0) i\sigma^{\mu\nu}\psi (0)$,
\ba\label{Eq:form-factors}
	\int\di x\,        H_T^q(x,\xi,t)&=&        F_T^q(t),\;\;\nonumber\\
	\int\di x\,\tilde{H}_T^q(x,\xi,t)&=&\tilde{F}_T^q(t),\;\;\nonumber\\
	\int\di x\,        E_T^q(x,\xi,t)&=&        E_T^q(t),\;\;\nonumber\\
	\int\di x\,\tilde{E}_T^q(x,\xi,t)&=& 0              \,. 
\ea
The vanishing of the first moment of $\tilde{E}_T^q$ is a consequence of the
antisymmetry in $\xi$, Eq.~(\ref{Eq:xi-dependence}); its higher moments are non-zero. 
More generally, the higher $x$--moments of the chiral-odd GPDs are polynomials in $\xi$
(generalized tensor form factors). 

In the limit of zero momentum transfer (forward limit) the chiral-odd GPD $H_T^q$ 
reduces to the transversity PDF $h_1^q (x)$,
\begin{equation}
H_T^q (x, \xi = 0, t = 0) \;\; = \;\; h_1^q (x) .
\label{H_T_forward}
\end{equation}
Its first moment is known as the nucleon's tensor charge. Because the local tensor 
operator is not a conserved current, the tensor charge is scale--dependent and
cannot directly be related to low--energy properties of the nucleon.

%
%
\begin{figure}[t]
\parbox[c]{.38\textwidth}
{\includegraphics[width=.35\textwidth]{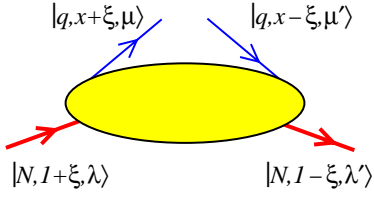}}
\hspace{.1\textwidth}
\parbox[c]{.5\textwidth}{\caption{\label{Fig-1:helicity-amplitude}
Representation of GPDs in the region $\xi < x < 1$ as nucleon--quark helicity amplitudes. 
In the nucleon and quark states (denoted as $N, q$) the second label denotes the fraction of the
light-cone plus momentum $P^+$ carried by the particle, and the third label denotes the 
light-cone helicity.}}	 
\end{figure}
%
%
The partonic interpretation of the GPDs is described in Refs.~\cite{Goeke:2001tz,
Diehl:2003ny,Belitsky:2005qn,Boffi:2007yc}. Our study of the large--$N_c$ limit of 
the chiral-odd GPDs relies essentially on the representation 
of the GPDs as partonic helicity amplitudes \cite{Diehl:2000xz}. 
This representation most naturally
appears in the region $\xi < x < 1$, where the GPDs describe the amplitude
for the ``emission'' by the nucleon of a quark with plus momentum fraction $x + \xi$ 
and subsequent ``absorption'' of a quark with $x - \xi$ 
(see Fig.~\ref{Fig-1:helicity-amplitude}). (In the region $-1 < x < -\xi$ the GPDs describe 
the emission and absorption of an antiquark, while in $-\xi < x <\xi$ they describe 
the emission of a quark--antiquark pair by the nucleon. We do not need to consider these 
regions explicitly in the subsequent arguments, as the $N_c$--scaling of the GPDs is 
uniform in $x$.) The initial and final nucleon are described by light-cone helicity
spinors, which are obtained from rest-frame spinors (polarized along the $z$--direction)
through a longitudinal and transverse boost; these spinors transform in a simple manner
under Lorentz transformations and have a clear connection to polarization states in 
the rest frame. The helicity amplitudes are then defined as
\be
	\!A^q_{\lambda^\prime\mu^\prime,\lambda\mu} = 
	\int\frac{\di z^-}{2\pi}\,e^{ixP^+z^-}  
	\la N(P^\prime,\lambda^\prime)|{\cal O}^q_{\mu^\prime\mu}
	|N(p,\lambda)\ra\biggl|_{{z^+=0},\;{\vec{z}_T=0}\,} ,
\ee
where $\lambda (\lambda')$ are the light-front helicity of the initial (final) nucleon
and $\mu (\mu')$ those of the initial (final) quark. Without loss of generality one may 
choose a frame in which the nucleon 3-momenta $\vec{p}$, $\vec{p}^{\,\prime}$ are in 
the x-z-plane. The chiral-odd light--ray operator associated with 
quark helicity flip is given by \cite{Diehl:2001pm}
\be
	{\cal O}^q_{+-} = 
	\psi_q(-\tfrac{z}{2})\;\frac{i}{4}\sigma^{+1}(1-\gamma_5)
	\;\psi_q(\tfrac{z}{2})\, .
\ee
The helicity amplitudes are related to chiral odd GPDs as 
\begin{subequations}
\ba\label{Eq:amplitudes-I}
&&	A^q_{++,+-}=\delta_k
	\biggl(\tilde{H}_T^q+\frac{1-\xi}{2}(E_T^q+\tilde{E}_T^q)\biggr) , \\
&&	A^q_{-+,--}=\delta_k
	\biggl(\tilde{H}_T^q+\frac{1+\xi}{2}(E_T^q-\tilde{E}_T^q)\biggr) , \\
&&	A^q_{++,--}=\sqrt{1-\xi^2}
	\biggl(H_T^q+\delta_k^2\tilde{H}_T^q
	-\frac{\xi^2}{1-\xi^2}E_T^q
	+\frac{\xi}{1-\xi^2}\tilde{E}_T^q\biggr) , \\
&&	A^q_{-+,+-}=\sqrt{1-\xi^2}\,
	\delta_k^2\;\tilde{H}_T^q\,, \phantom{\frac11} 
\label{Eq:amplitudes-I-end}\ea
\end{subequations}
where the ``kinematic'' prefactor $\delta_k$ is defined as
\be\label{Eq:delta-k}
	\delta_k = {\rm sign}\biggl(P^+\Delta^1-\Delta^+P^1\biggr)
	\frac{\sqrt{t_0-t}}{2M_N}\;,\;\;\;
	-t_0 = \frac{4M_N^2\xi^2}{1-\xi^2}\; ,
\ee
in which $-t_0$ is the minimal value of $-t$ for the given value of $\xi$.

In the $1/N_c$ expansion it is convenient to work with linear combinations of the 
helicity amplitudes Eqs.~(\ref{Eq:amplitudes-I}-\ref{Eq:amplitudes-I-end}) 
that have a simple interpretation 
in terms of spin transitions in the nucleon rest frame. We define them as
\begin{subequations}
\begin{align}
&	A_0^q = \frac12\biggl(A^q_{++,+-} + A^q_{-+,--}\biggr) 
        \propto \delta^{}_{\lambda\lambda^\prime} \; ,\label{Eq:amplitudes-IIa}\\
&	A_1^q = \frac12\biggl(A^q_{++,--} + A^q_{-+,+-}\biggr) 
	\propto \sigma^1_{\lambda\lambda^\prime} \; ,\label{Eq:amplitudes-IIb}\\
&	A_2^q = \frac12\biggl(A^q_{++,--} - A^q_{-+,+-}\biggr) 
	\propto \sigma^2_{\lambda\lambda^\prime} \; ,\label{Eq:amplitudes-IIc}\\
&	A_3^q = \frac12\biggl(A^q_{++,+-} - A^q_{-+,--}\biggr)  
	\propto \sigma^3_{\lambda\lambda^\prime} \; ,\label{Eq:amplitudes-IId}
\end{align}
\end{subequations}
where $\sigma^a_{\lambda\lambda^\prime}$ are the Pauli spin matrices. In the 
nucleon rest frame, these combinations describe spin transitions in which the
matrix element of the spin operator has components along the $x, y$ and $z$ axes. 
\section{$1/N_c$ expansion of chiral-odd GPDs}
\label{Sec-3:large-Nc}
In the large-$N_c$ the nucleon mass scales as $M_N \sim N_c$, while the nucleon size
remains stable, $R \sim N_c^0$. The $1/N_c$ expansion of GPDs is performed in a class
of frames where the initial and final nucleon move with 3--momenta 
$p^k, \,p^{\prime k} \sim N_c^0 \; (k = 1, 2, 3)$
and have energies $p^0, \, p^{\prime 0} = M_N + O(1/N_c)$, which implies an 
energy and momentum transfer $\Delta^0 \sim N_c^{-1}, \, \Delta^k \sim N_c^0$, 
and thus
\begin{equation}
\Delta^i \; \sim \; N_c^0 \hspace{2em} (i = 1, 2), \hspace{2em} \xi \; \sim \; N_c^{-1}, 
\hspace{2em} |t| \; \sim \; N_c^0 . 
\end{equation}
In the partonic variable $x$ one considers the parametric region
\begin{equation}
x \; \sim \; N_c^{-1} ,
\label{nc_scaling_x}
\end{equation}
corresponding to non-exceptional longitudinal momenta of the quarks and antiquarks 
relative to the slowly moving nucleon, $x M_N \sim R^{-1} \sim  N_c^0$. Likewise,
it is assumed that the normalization scale of the light-ray operator is $\sim N_c^0$, 
so that the typical quark transverse momenta are $\sim N_c^0$. Equation~(\ref{nc_scaling_x})
corresponds to the intuitive picture of a nucleon consisting of $N_c$ ``valence'' quarks
and a ``sea'' of $O(N_c)$ quark--antiquark pairs, with each quark/antiquark carrying
on average a fraction $\sim 1/N_c$ of the nucleon momentum. Altogether, the 
$N_c$--scaling relations for GPDs can then expressed in the form
\begin{equation}
{\rm GPD}(x, \xi, t) \;\; \sim \;\; N_c^k \; \times \; \textrm{Function}(N_c x, N_c\xi, t) ,
\label{nc_scaling_gpd_generic}
\end{equation}
where the scaling exponent $k$ depends on the spin--flavor structure and can be 
established on general grounds, while the scaling function on the right-hand-side
does not explicitly depend on $N_c$ and can only be determined in specific 
dynamical models.

A general method for the $1/N_c$ expansion of nucleon matrix elements of quark bilinear
operators has been given in Ref.~\cite{Goeke:2001tz}. 
It uses the fact that the large--$N_c$ nucleon is described by a classical field with a
certain spin--isospin symmetry, and that the nucleon states of definite spin, isospin, 
and momentum are obtained through quantization of the (iso--) rotational and translational 
zero modes \cite{Witten:1983tx}. 
In this classical picture the nucleon matrix element of the quark bilinear 
operator is obtained by taking the expectation value of the operator in the static 
classical field, allowing for collective (iso--) rotations, and computing the
transition matrix element between collective wave functions corresponding to the
desired nucleon spin--isospin and momentum states. While the expectation value of the 
operator in the classical field cannot be calculated from first principles and
remains unknown, its symmetry properties are unambiguously determined by the
spin--isospin symmetry of the classical field and the quantum numbers of the
quark bilinear operator. Altogether, this allows one to establish the
$N_c$--scaling of the different spin--isospin components of the matrix element
without calculating the coefficients accompanying the powers of $1/N_c$.

We use the method of Ref.~\cite{Goeke:2001tz} to establish the $N_c$--scaling of the 
nucleon--quark helicity amplitudes and the chiral-odd GPDs. It is natural to work with 
the combinations Eqs.~(\ref{Eq:amplitudes-IIa}--\ref{Eq:amplitudes-IId}),
which correspond 
to definite spin transitions in the frame where the nucleons are moving slowly, and which
exhibit homogeneous scaling in $1/N_c$. When expressing the helicity amplitudes in terms of
the GPDs, we use that in the large--$N_c$ limit the kinematic factor 
$\delta_k\sim N_c^{-1}$ in (\ref{Eq:delta-k}) and the variable $\xi$ become
\be\label{Eq:delta-k-in-large-Nc}
	\delta_k=\frac{\Delta^1}{2 M_N}\;, \;\;\;
	\xi = -\;\frac{\Delta^3}{2 M_N}\;. \;\;\;
\ee
In this way we obtain in leading order of the $1/N_c$ expansion the relations
\begin{subequations}
\begin{align}
	N_c^2\sim A_0^{u+d} = &
	\frac{\delta_k}{2}\biggl(E_T^{u+d}+2\tilde{H}_T^{u+d}\biggr) ,
	\label{Eq:work2-Large-0} \\
	N_c^2\sim A_1^{u-d} = &
	\;\frac12\biggl(H_T^{u-d}+\xi\tilde{E}_T^{u-d}\biggr) ,
	\label{Eq:work2-Large-1} \\
        N_c^2\sim A_2^{u-d} = &
	\;\frac12\biggl(H_T^{u-d}+\xi\tilde{E}_T^{u-d}\biggr) ,
	\label{Eq:work2-Large-2} \\
	N_c^2\sim A_3^{u-d} = & 
	\frac{\delta_k}{2}\biggl(\tilde{E}_T^{u-d}\biggr) ,
	\label{Eq:work2-Large-3}
\end{align}
\end{subequations}
which are to be understood in the sense of Eq.~(\ref{nc_scaling_gpd_generic}), and in
which we indicate only the scaling exponent. In subleading order of the $1/N_c$ expansion,
\begin{subequations}
\begin{align}
	N_c\sim A_0^{u-d} = & \frac{\delta_k}{2}\biggl(
	E_T^{u-d}-\xi\tilde{E}_T^{u-d}+2\tilde{H}_T^{u-d}\biggr) ,
	\label{Eq:work2-small-0}\\
	N_c\sim A_1^{u+d} = &
	\;\frac12\biggl(H_T^{u+d}-\xi^2E_T^{u+d}+\xi\tilde{E}_T^{u+d}
	+2\delta_k^2\tilde{H}_T^{u+d}\biggr) ,
	\label{Eq:work2-small-1}\\
      	N_c\sim A_2^{u+d} = & \;\frac12\biggl(
	H_T^{u+d}-\xi^2E_T^{u+d}+\xi\tilde{E}_T^{u+d}\biggr) ,
	\label{Eq:work2-small-2}\\
	N_c\sim A_3^{u+d} = &
	\frac{\delta_k}{2}\biggl(\tilde{E}_T^{u+d}-\xi E_T^{u+d}\biggr) .
	\label{Eq:work2-small-3}
\end{align}
\end{subequations}
From Eqs.~(\ref{Eq:work2-Large-0}---\ref{Eq:work2-small-3}) we can read off 
the large-$N_c$ flavor structure of the GPDs, namely
\ba
\label{Eq:H_T}{H}_T^{u-d} \sim N_c^2 \;, &&       {H}_T^{u+d} \sim N_c   \;,\\
	      {E}_T^{u+d} \sim N_c^3 \;, &&       {E}_T^{u-d} \sim N_c^2 \;,\\
	\tilde{H}_T^{u+d} \sim N_c^3 \;, && \tilde{H}_T^{u-d} \sim N_c^2 \;,\\
	\tilde{E}_T^{u-d} \sim N_c^3 \;, && \tilde{E}_T^{u+d} \sim N_c^2 \;.
\ea
These relations generalize previous results for the $N_c$--scaling of the nucleon's
quark transversity PDF $h_1^{q}$ [obtained as the forward limit of $H_T^q$, 
Eq.~(\ref{H_T_forward})] \cite{Pobylitsa:1996rs,Schweitzer:2001sr,Pobylitsa:2003ty}
and the tensor form factors [obtained as the first $x$--moment, 
Eq.~(\ref{Eq:form-factors})] \cite{Ledwig:2010tu,Ledwig:2011qw}.

The scaling relations Eq.~(\ref{Eq:H_T}) reveal several interesting properties of
the chiral-odd GPDs. First, one sees that in $H_T$ and $\tilde E_T$ the
flavor--nonsinglet component is leading and the flavor--singlet one subleading,
while in $E_T$ and $\tilde H_T$ the order is opposite. 

Second, one notices that the amplitudes $A_1^{u-d}$ and $A_2^{u-d}$ 
are degenerate in leading order of the $1/N_c$ expansion; see
Eqs.~(\ref{Eq:work2-Large-1},~\ref{Eq:work2-Large-2}). This happens because
the mean--field picture implied by the large--$N_c$ limit makes no distinction 
between different transverse polarization directions in leading order of the 
$1/N_c$ expansion. The equality $A_1^{u-d} = A_2^{u-d}$ implies that the nucleon--quark
double helicity--flip amplitude vanishes, $A^{u-d}_{-+,+-} = 0$,
i.e., the amplitude in which the quark has helicity \textit{opposite} to the nucleon 
in the initial state and \textit{both} helicities are flipped in the final state.
If we view the nucleon as a system
composed of the active quark and a ``remnant,'' the double--flip transition involves 
a change of the relative orbital angular momentum by two units, 
$\Delta L = 2$ \cite{Diehl:2001pm}. 
Among all the nucleon--quark helicity amplitudes (including also the chiral-even ones) 
this is the only amplitude that is (i) double--flip, and (ii) vanishes exactly 
in the leading order of the large-$N_c$ limit.
It merits further study whether this observation could be explained more directly 
in light of the mechanical picture implied by the large-$N_c$ limit.

Third, one notices in Eqs.~(\ref{Eq:work2-Large-0}---\ref{Eq:work2-small-3})
that the combination $E_T^{q} + 2\tilde{H}_T^{q} \equiv \bar E_T^q$ emerges naturally 
from the large-$N_c$ expansion of the chiral-odd GPDs. The differences between the 
individual GPDs $E_T^{q}$ and $\tilde{H}_T^{q}$ are suppressed in $1/N_c$.
Interestingly, this combination appears also in the amplitude for pseudoscalar
meson production \cite{Goloskokov:2011rd}, 
meaning that the virtual photon cannot ``distinguish'' between
$E_T^{q}$ and $\tilde{H}_T^{q}$.

\section{Comparison with pseudoscalar meson production data and outlook}
\label{Sec-5:comparison}
It is interesting to compare our results with preliminary data from the JLab CLAS 
exclusive pseudoscalar meson production experiments 
\cite{Bedlinskiy:2012be,Kubarovsky:2014spin} 
(cf.\ comments in Sec.~\ref{Sec-1:introduction}).
Analysis of the azimuthal--angle dependent response functions shows that 
$|\sigma_{LT}| \ll |\sigma_{TT}|$, which indicates dominance of the twist-3 amplitudes, 
involving the chiral-odd GPDs $H_T^q$ and $\bar E_T = E_T^{q} + 2\tilde{H}_T^{q}$,
over the twist--2 amplitudes involving the chiral-even GPD $\tilde E^q$. A preliminary 
flavor decomposition was performed assuming dominance of the twist--3 amplitudes and
combining the data on $\pi^0$ and $\eta$ production, in which the $u$ and $d$ quark 
GPDs enter with different relative weight. Results show opposite sign of the
exclusive amplitudes $\langle H_T^u \rangle$ and $\langle H_T^d \rangle$, which is 
consistent with the leading appearance of the flavor-nonsinglet $H_T^{u - d}$ in the
$1/N_c$ expansion. (Here $\langle\ldots \rangle$ denotes the integral over $x$ of 
the GPD, weighted with the meson wave function, hard process amplitude, and Sudakov 
form factor \cite{Goloskokov:2011rd}.) The results also suggest same sign of
$\langle \bar E_T^u \rangle$ and $\langle \bar E_T^d \rangle$, which is again
consistent with the leading appearance of the flavor-singlets $E_T^{u + d}$ and 
$\tilde H_T^{u + d}$ in the $1/N_c$ expansion. These findings should be interpreted
with several caveats: (a) the errors in the experimental extraction of
$\langle H_T^q \rangle$ and $\langle \bar E_T^q \rangle$ are substantial; 
(b) the $1/N_c$ expansion predicts only the scaling behavior, not the absolute
magnitude of the individual flavor combinations, cf.\ Eq.~(\ref{nc_scaling_gpd_generic}).

It is encouraging that the flavor structure of the amplitudes extracted from the 
$\pi^0$ and $\eta$ electroproduction data is consistent with the pattern predicted by
the $1/N_c$ expansion. Our findings further support the idea that pseudoscalar meson
production at $x_B \gtrsim 0.1$ and $Q^2 \sim \textrm{few GeV}^2$ is 
governed by the twist-3 mechanism involving the chiral-odd GPDs. It would be interesting
to calculate the chiral-odd GPDs in dynamical models that consistently implement the
$N_c$--scaling, such as the chiral quark--soliton model. Such calculations would allow
one to calculate also the scaling functions in the large--$N_c$ relations,
Eq.~(\ref{nc_scaling_gpd_generic}), and supplement the scaling studies with
dynamical information.

\begin{acknowledgments}
We thank V.~Kubarovsky for discussions on the preliminary JLab CLAS pseudoscalar 
meson production data. In our study we greatly benefited from discussions with D.~Diakonov, 
V.~Petrov, P.~Pobylitsa, and M.~Polyakov during earlier joint work. 
P.S.\ is supported by NSF under Contract No.~1406298.
{\it Notice:} Authored by Jefferson Science Associates, LLC
under U.S.~DOE Contract No.~DE-AC05-06OR23177.
The U.S.~Government retains a non-exclusive, paid-up,
irrevocable, world-wide license to publish or reproduce
this manuscript for U.S.~Government purposes.
\end{acknowledgments}

\end{document}